\documentclass[a4paper,amssymb,amsmath,pra,onecolumn,floatfix]{revtex4}
\usepackage{graphicx}
\usepackage{lastpage}
\usepackage[normalem]{ulem}
\usepackage{url,lineno}
\usepackage{amssymb}
\usepackage{amsmath}
\usepackage{breqn}
\usepackage{color}
\usepackage{xparse}
\usepackage{dcolumn}
\usepackage{bm}
\usepackage{gensymb}
\usepackage{amsmath}
\usepackage{tabularx}
\usepackage{pbox}
\usepackage{caption}

\newcommand{\ket}[1]{|#1 \rangle}
\newcommand{\bra}[1]{\langle #1|}

\newcommand{\ketbra}[2]{|#1\rangle\langle #2|}

\newcommand\T{\rule{0pt}{2.6ex}}       
\newcommand\Tt{\rule{0pt}{3.6ex}}  
\newcommand\Ttt{\rule{0pt}{5.2ex}}  
\newcommand\Tttt{\rule{0pt}{9.6ex}}  
\newcommand\B{\rule[-1.2ex]{0pt}{0pt}} 

\usepackage[normalem]{ulem} 
\definecolor{orange}{rgb}{0.7,0.2,0}
\definecolor{darkgreen}{rgb}{0,0.4,0}

\def\Authors{Caitlin Batey$^1$, Jan Jeske$^1$, and Andrew D.~Greentree,$^{1,2}$}

\def\corrAddress{Chemical and Quantum Physics, School of Applied Sciences, RMIT University, Melbourne 3001, Australia}

\begin{document}


\title{Dark state adiabatic passage with branched networks and high-spin systems: spin separation and entanglement}


\author{\Authors}
\address{\corrAddress}



\begin{abstract}
Adiabatic methods are potentially important for quantum information protocols because of their robustness against many sources of technical and fundamental noise.  They are particularly useful for quantum transport, and in some cases elementary quantum gates.  Here we explore the extension of a particular protocol, dark state adiabatic passage, where a spin state is transported across a branched network of initialised spins, comprising one `input' spin, and multiple leaf spins.  We find that maximal entanglement is generated in systems of spin-half particles, or where the system is limited to one excitation.  
\end{abstract}

\maketitle


\section{Introduction}
Techniques of adiabatic passage are of interest from fundamental viewpoints, and also for their robustness against many sources of technical noise \cite{KTS2007}.  The canonical example is stimulated Raman adiabatic passage, STIRAP, \cite{GRB+1988}, where an electron is transported between two meta-stable states, by applying two laser pulses that couple the meta-stable states to an excited state, or occasionally a continuum \cite{PYH2005,DSH+2009}.  By applying the laser pulses adiabatically in the so-called counter-intuitive ordering, where the unpopulated transition is coupled before the laser that addresses the particle to be transferred, the transported particle can move between the two states without ever populating the excited state.  In this way the protocol shows robustness against spontaneous emission, and because the population at any time only goes like the ratio of the Rabi frequencies of the two fields, STIRAP is also robust against fluctuations in the total laser intensity when the fields come from the same source.

A conceptual breakthrough occurred in 2002 when Brandes and Vorrath \cite{BV2002} reported a method to use STIRAP to transfer electrons between wells in a double quantum dot.  This was significant as it was the first time that the use of \emph{engineered}, as opposed to naturally occurring, systems was considered.  Following spatial STIRAP, schemes have been proposed where only the spatial tunnelling interaction is varied to effect the counter-intuitive pulse sequence, and these are sometimes termed coherent tunnelling adiabatic passage, CTAP.  Eckert \textit{et al.} considered atomic transport through optical potentials \cite{ECB+2004}, Siewert and Brandes proposed transport in superconducting networks \cite{SB2004}, and Greentree \textit{et al.} proposed electronic transport in quantum-dot and phosphorus in silicon systems \cite{GCH+2004}.  Later proposals included Bose-Einstein condensates \cite{GKW2006,RCP+2008}, optical waveguides \cite{P2006,LDO+2007}, Bose-Hubbard systems \cite{BRG+2012}, sonic systems \cite{MMA2014}, polarisation \cite{DRK2015}, and spin chains \cite{OEO+2007,OSF+2013,GK2014}.  Spatial adiabatic passage in spin chains is usually termed dark state adiabatic passage, DSAP, and is the subject of this work.

The combination of designed Hilbert spaces with adiabatic processes leads to interesting new protocols and opportunities.  In particular, adiabatic passage enables quantum gates via the method outlined by Unanyan, Shore and Bergman \cite{USB1999,KR2002,DGH2007,HNM+2015}; interferometers \cite{JG2010}, interaction-free measurements \cite{HGH2011},  and robust splitters \cite{DOS+2009,CCR+2012,CKR+2012}.

\begin{figure}
\centering
    \includegraphics[width=0.5\linewidth]{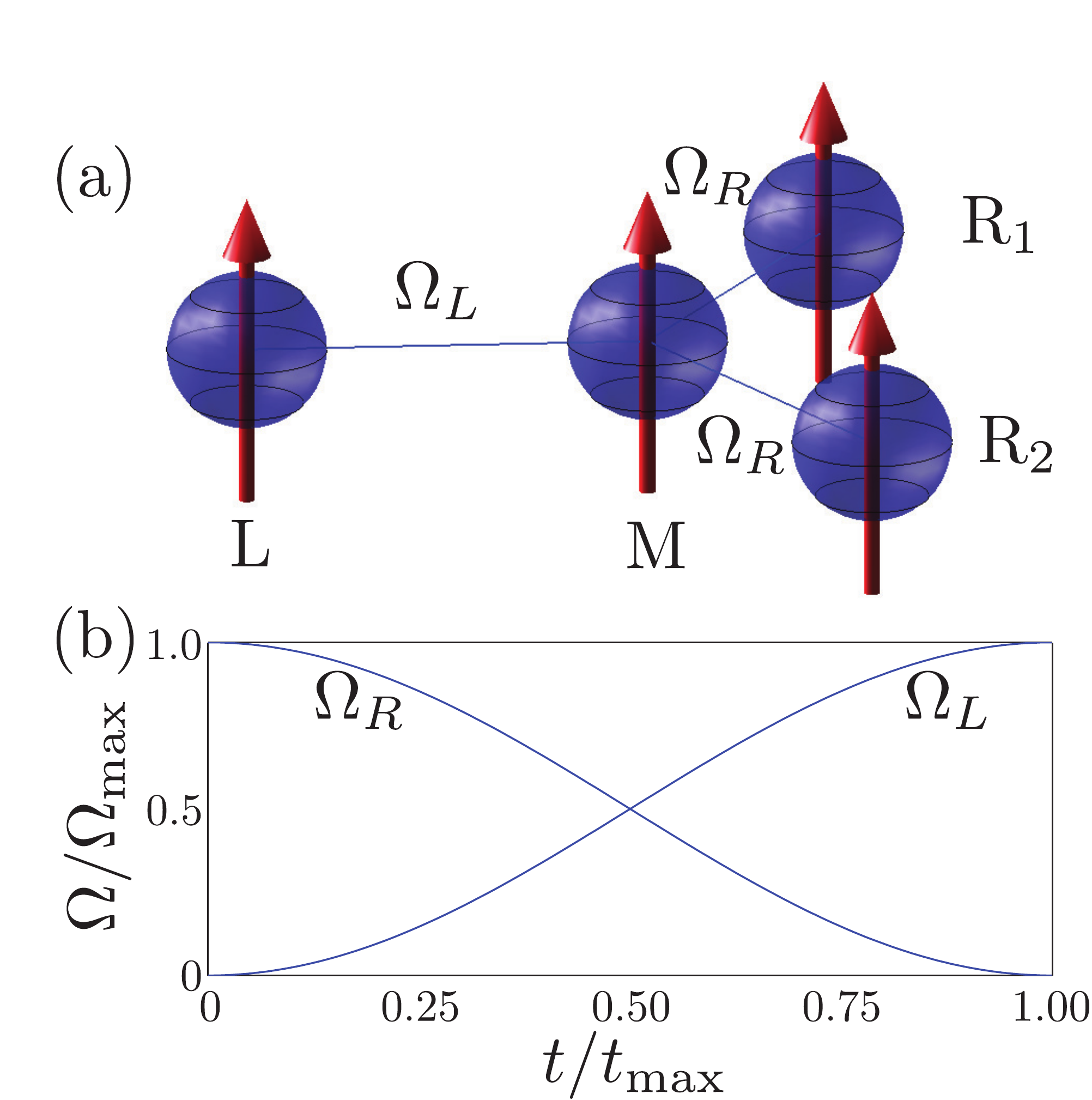}
    \caption{(a) Schematic showing a network for four spin-one particles, with starting node, $L$, central node $M$ and two leaf nodes $R_1$ and $R_2$.  Spin-spin couplings are indicated by the solid lines connecting the spins. (b) Dark state adiabatic passage is achieved by utilising the counter-intuitive pulse sequence.  To transport spin information from $L$ to $R_1$ and $R_2$ requires $\Omega_R(0)\gg\Omega_L(0)$ and $\Omega_L(t_{\max}) \gg \Omega_R(t_{\max})$.  The exact shape of the pulse is of little importance provided that adiabaticity is preserved and here we have chosen squared sinusoidal coupling for simplicity.}
    \label{fig:Fig1}
\end{figure}

Here we consider the combination of DSAP with branched spin networks to explore how entanglement is generated between the leaf nodes, i.e.~end points of the network.  We also note that adiabatic passage on lattice networks has been considered \cite{L2014}. Our aim can be understood with respect to Fig.~\ref{fig:Fig1} (a), which shows a simple network where all of the spins are initialised in their lowest spin state, except for the `initial' spin, which is set to some particular state.  We perform the counter-intuitive pulse sequence (described explicitly below) with an XX + YY + ZZ Hamiltonian, and explore how the initial spin state is distributed between the leaf nodes.  In its simplest form, this case has been considered as a means to create a superposition of a single particle amongst the leaf nodes \cite{HGH2011,GDH2006,RV2012}, and this case is isomorphic to the spin-half case discussed below.   We extend this work by exploring networks of higher-spin systems and multiple leaf nodes.  We observe more complicated behaviour, although such behaviour does not appear to lead to entangled states superior to the spin-half case.

This paper is organised as follows: We first describe the Hamiltonian for DSAP with high-spin systems.  We then show results for spin-half, spin-one and spin-three half networks.  In all cases we show the adiabatic evolution and quantify the final entanglement generated.  Finally we compare the protocols and offer perspectives.

\section{Hamiltonian}

In its most general form, we consider the adiabatic transport of spin information from an initial spin, $L$, to some entangled state of final (or leaf) spins, $R_1$ to $R_N$, via a middle spin, denoted $M$, which takes the analogous role of the unpopulated excited state in STIRAP.  We assume a uniform, time invariant magnetic field $B$ aligned with the $Z$ direction which energetically separates manifolds of different numbers of excitations. We assume time-varying inter-site couplings between the spins and the middle spin only, where the $L-M$ coupling strength is $\Omega_L(t)$ and the $M-R_j$ coupling strength is $\Omega_{R_j}$.  The Hamiltonian for our network can be expressed as 

\begin{align}
H = \sum_{i} B J^Z_{i}  +  \left[\Omega_L(t) J_L^+ J_M^- + \Omega_R(t) \sum_{j} J_M^+ J_{R_j}^- + \text{H.c.}\right] + \alpha(J_L^Z J_M^Z + \sum_j J_M^Z J_{R_j}^Z),
\end{align}

\noindent where the index $i$ ranges across all spins $\{L, M, R_1 ... R_N\}$, and the index $j$ ranges across all leaf spins $R_1$ to $R_N$. $J_z$, $J^+$ and $J^-$ are the Z-projection of the spin operator, spin raising and spin lowering operators respectively. The Hamiltonian preserves the number of excitations meaning that subspaces of different numbers of excitation can be treated separately in the adiabatic passage. $\alpha$ is the coefficient of the always-on ZZ coupling and is to prevent occupation of spin $M$, and operates in a similar fashion  to the role of central state detuning in Ref.~\cite{BRG+2012}.  The value of $\alpha$ was set to $1\times 10^{-2} B$ for all systems.

DSAP of the spin state from the initial to the final site(s), utilises the counter-intuitive pulse ordering, i.e. the $\Omega_{R_i}$ coupling should be high before $\Omega_L$.  For our numerical results, for simplicity we have chosen:

\begin{align}
\Omega_L(t) = \Omega_{\max} \sin\left(\frac{\pi t}{2t_{\max}}\right), \qquad
\Omega_{R_j}(t) = \Omega_{\max} \cos\left(\frac{\pi t}{2t_{\max}}\right), \label{eq:CIPulse}
\end{align}

\noindent where $t_{\max}$ is the total time for the transport, and $0 \leq t < t_{\max}$ is the time.  This is shown in Fig.~\ref{fig:Fig1}(b).  In keeping with other adiabatic passage protocols, the exact form of the $\Omega$ is relatively unimportant providing that the counter intuitive ordering is maintained, i.e. $\Omega_L(t) \ll \Omega_{R_j}(t)$ as $t\rightarrow 0$, $\Omega_L(t) \gg \Omega_{R_j}(t)$ as $t\rightarrow t_{\max}$, and the variation in the $\Omega$ satisfy the adiabaticity criterion, so that for every pair of eigenstates $\ket{\psi}$, $\ket{\phi}$ with respective eigenenergies $E_{\ket{\psi}}$ and $E_{\ket{\phi}}$

\begin{align}
\bra{\phi}\partial_t H \ket{\psi} \ll |E_{\ket{\phi}} - E_{\ket{\psi}}|.
\end{align}

\noindent Although our calculations are performed explicitly as a function of time and without assuming adiabaticity, all of the results we show below are in the adiabatic limit.

While the Hamiltonian and pulse ordering can allow for adiabatic passage of the state in site L to the recipient spins R$_j$, it is important to stress that DSAP is also dependent on the choice of the initial state of all spins other than L. For simplicity, we will assume a state with all spins $M$ and $R_j$ prepared in their lowest spin projection with respect to the $z$ axis.  Another enabling  state has all spins in the highest spin projections and yields completely symmetric results. It is possible that there are other states that will permit DSAP, as was the case for direct transport along a one-dimensional chain \cite{GK2014}, although we will not explore such possibilities here.

We quantify the DSAP by two methods.  First is to identify the final states for various network configurations.  Secondly, we examine the entanglement that is generated between the $R_j$ nodes.  Because we are dealing with multi-partite systems of varying dimensionality, there is no unique way to quantify entanglement.  For simplicity, we choose to use the entanglement of formation \cite{W2001,WC2001}, which is an entanglement monotone for pure states where 0 corresponds to no entanglement and 1 to maximal entanglement for that system.  The entanglement of formation for a state $\ket{\psi}$ is defined

\begin{align}
E(\ket{\psi}) \equiv -\mathrm{Tr}(\rho_{R_1} \log_{2} \rho_{R_1}) = -\mathrm{Tr}(\rho_{R_2} \log_{2} \rho_{R_2})
\end{align}
\noindent where 
			$\rho_{R_1}$ is the partial trace of $\ketbra{\psi}{\psi}$ over subsystem $R_2$ and
			$\rho_{R_2}$ is the partial trace of $\ketbra{\psi}{\psi}$ over subsystem $R_1$.
In the configurations with more than two recipients we quantify the entanglement of formation as being between one arbitrary leaf spin and all other recipient leaf spins.

\section{Spin-half networks}
The case of spin-half networks with one excitation is formally equivalent to the case of multi-recipient adiabatic passage \cite{GDH2006,RV2012} and so we consider it here purely for the purposes of review and to compare with our other results.

\begin{figure}[tb!]
\centering
    \includegraphics[width=0.5\linewidth]{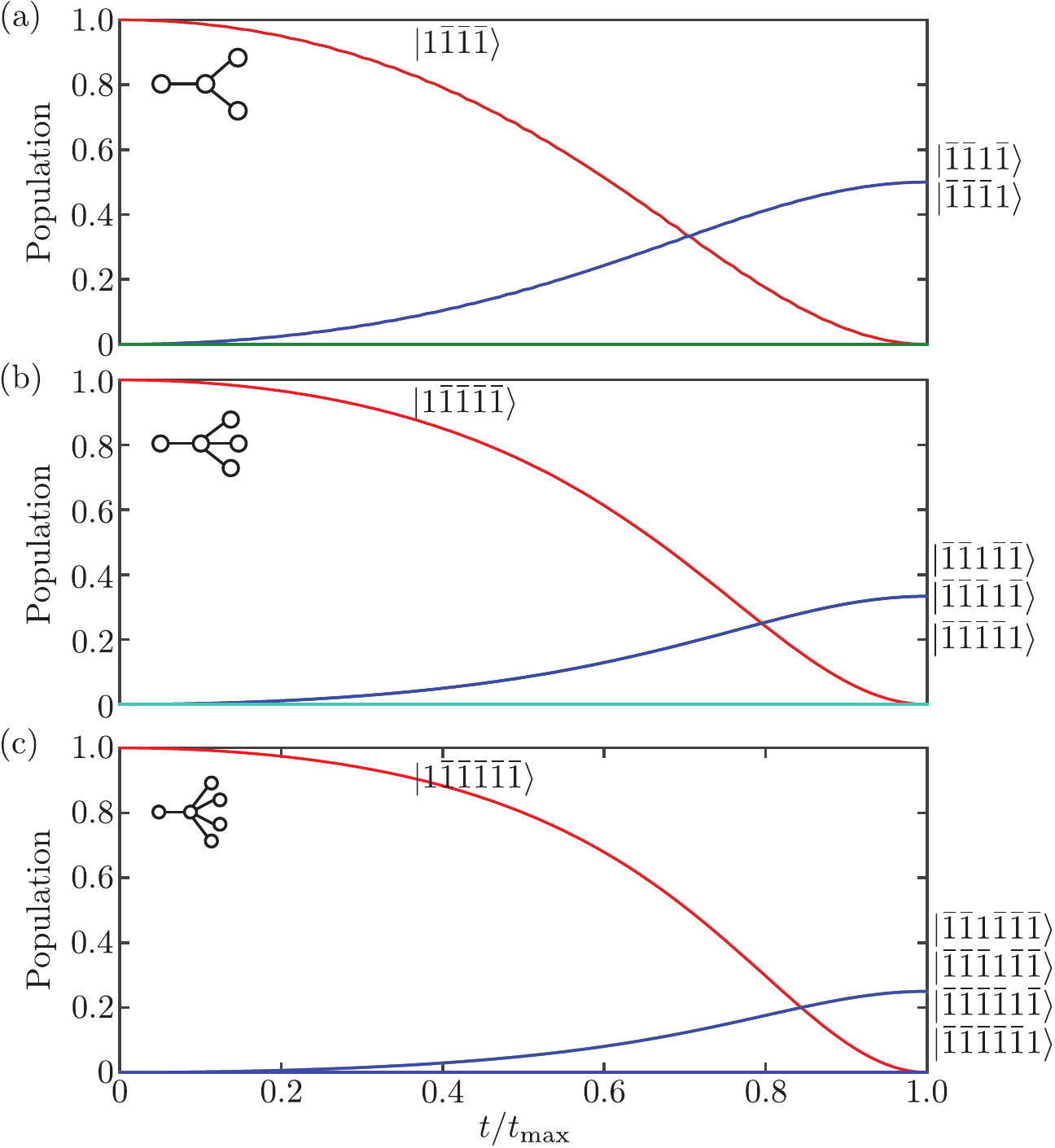}
    \caption{Population evolution for spin-half networks with (a) two leaf nodes, (b) three leaf nodes, and (c) four leaf nodes. The excitation begins at $L$  i.e. in the state $\ket{1\bar{1}...\bar{1}}$ and moves to an equally weighted superposition of all of the leaf nodes, which is a W state across the spins $R_i$. All initial and final states as well as the quantified entanglement related to each protocol is summarised in Table~\ref{table:nonlin}. }
    \label{fig spin half}
\end{figure}

We denote the spin projections of the spins with respect to their $J_z$ eigenvalue.   We define $\ket{1}$ as the state with spin projection parallel to the external magnetic field, i.e. with spin projection $\hbar/2$; and  $\ket{\bar{1}}$ as the state antiparallel to the external field, with spin projection $ -\hbar/2$. 
The middle and leaf (right) spins are assumed initialised in $\ket{\bar{1}}$, and we study the adiabatic passage for different initial (left) spin states.   The evolution is indicated in Fig.~\ref{fig spin half}.
The general form of the dark state, i.e.~the eigenstate which defines the adiabatic passage, for $n$ leaf nodes for the case considered above is

\begin{align}
\ket{D_0} = \frac{n\Omega_R \ket{1\bar{1}\bar{1}\bar{1}...\bar{1}} - \Omega_L \left( \ket{\bar{1}\bar{1}1\bar{1}...\bar{1}}  + \ket{\bar{1}\bar{1}\bar{1} 1 ... \bar{1}} + \cdots + \ket{\bar{1}\bar{1}\bar{1}...1}\right)}{\sqrt{n^2 \Omega_R^2 + n \Omega_L^2}}. \label{eq:D0Spin1/2}
\end{align}

\noindent Equation \eqref{eq:D0Spin1/2} shows that for the two leaf configuration, an initial excitation at $L$ ($\Omega_L=0$) is adiabatically transported to an equally weighted, maximally entangled state of the $R_j$ ($\Omega_R=0$).  As the number of leaves is increased, we see the formation of W-like states, which have a lower entanglement (as quantified by the entanglement of formation) with increasing $n$.  

\section{Spin-one networks}
Spin one networks are more complicated than spin-half, and are no longer isomorphic to the simple case of direct particle transport, although in certain cases a spin-one chain can be treated as a form of particle transport \cite{H83,AKL+87}.  Spin-one DSAP faithfully transmits the state of an arbitrary qutrit from one end of the chain to the single recipient spin at the other end of the chain \cite{GK2014}.  However, as we will show, the branched geometry maps the doubly-excited state partially into product states of two sites with one excitation each, rather than solely an entangled state of both excitations in one or the other site, thereby failing to preserve the integrity of the qutrit.  

As before, we define our basis states with respect to the Z projections of the individual spins. The states are $\ket{1}, \ket{0}, \ket{\bar{1}}$ corresponding to the spin projections  $\hbar$, 0 and $-\hbar$. As before, we initialise spins M and R$_i$ to $\ket{\bar{1}}$ (antiparallel to the Z field). In the spin-one system, transport from L to R$_j$ can be achieved for either one excitation ($\ket{0}$) or two excitations ($\ket{1}$). 

\begin{figure}[tb!]
\centering
    \includegraphics[width=\linewidth]{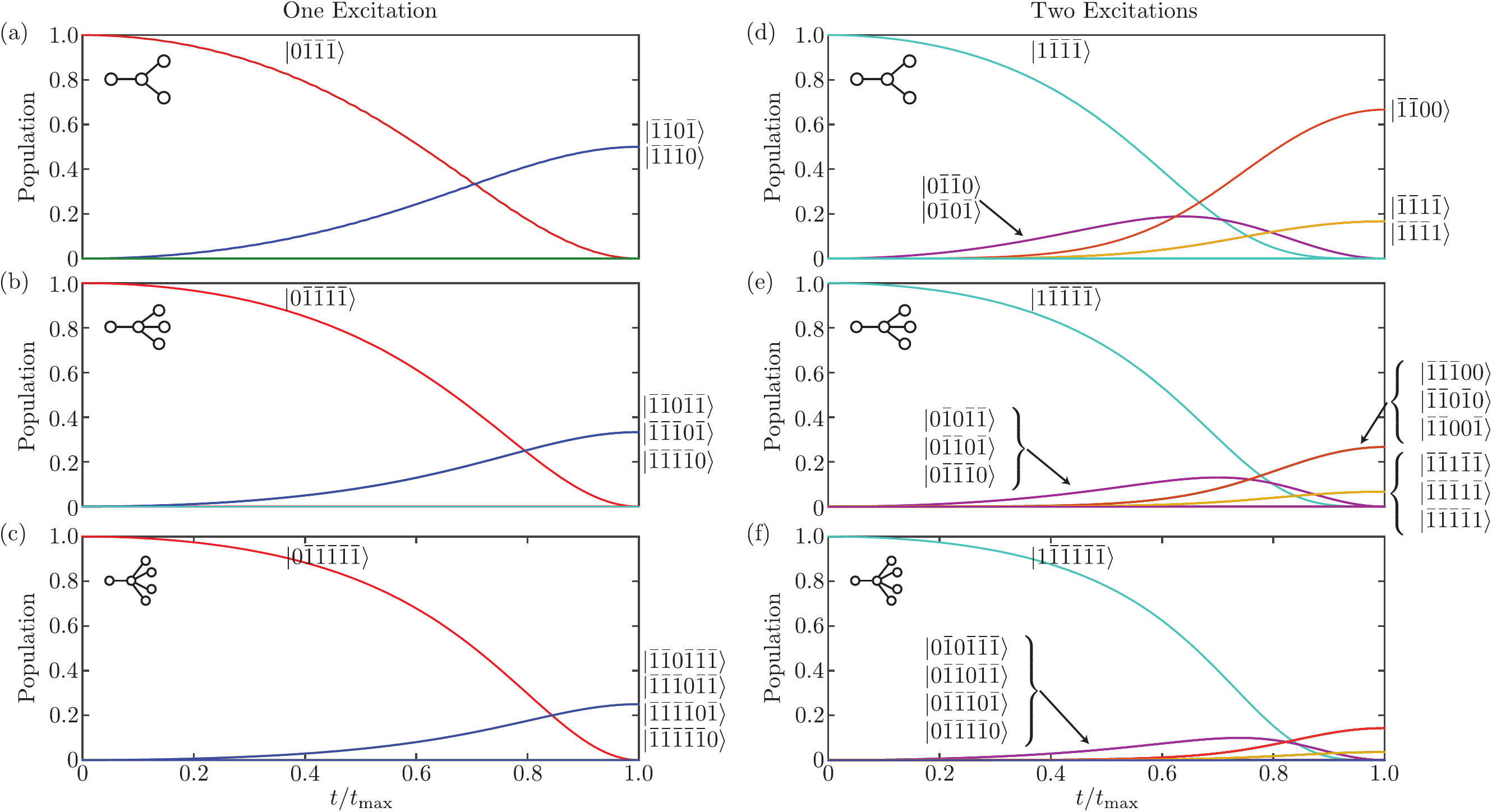}
    \caption{Spin-one networks for DSAP.  For the case of one excitation and (a) two leaf nodes, (b) three leaf nodes, and (c) four leaf nodes; we observe results equivalent to those seen for the spin-half networks. For two excitations and (d) two leaf nodes, (e) three leaf nodes, and (f) four leaf nodes, we observe evolution to a more complicated final state, via a transient, W-like state, as shown.  The sharing of the population between the states is determined by the egalitarian principle as discussed in the text.  Final states for two excitations and four nodes are not given here due to the large number of states involved, but is quantified, along with entanglement of formation, in Table~\ref{table:nonlin}. }
    \label{fig spin one}
\end{figure}

When limiting the system to one excitation, i.e. choosing the initial state of spin $L$ to be $\ket{0}$, the transport achieved in all configurations is the equivalent of the spin-half case and the entanglement of formation for both spin-one and spin-half is the same.  The evolution for three configurations, with two, three and four leaf nodes, is shown in Figs.~\ref{fig spin one}(a), (b) and (c).

In the two-excitation subspace for two leaf nodes, the evolution becomes more complicated due to the number of ways the excitations can be shared between the leaf nodes. Figure~\ref{fig spin one}(d) shows the transport with starting state $ \ket{1\bar{1}\bar{1}\bar{1}}$ with two leaf nodes. The counter intuitive pulse sequence ensures that all of the excitation is transported from $L$ to the $R_j$ without final occupation of $M$.  However the final state is complicated by the fact that there are more ways for the excitation to be shared between the $R_j$ than in the one excitation case.  The pertinent eigenstate for the DSAP is

\begin{align}
\ket{D_0} = \frac{2\Omega_R^2 \ket{1\bar{1}\bar{1}\bar{1}} + \Omega_L\Omega_R\left(\ket{0\bar{1}\bar{1}0} + \ket{0\bar{1}0\bar{1}}\right) + \Omega_L^2\left( 2\ket{\bar{1}\bar{1}00} +\ket{\bar{1}\bar{1}1\bar{1}} + \ket{\bar{1}\bar{1}\bar{1}1}\right)}{\sqrt{4\Omega_R^2 + 2\Omega_L^2\Omega_R^2 + 6\Omega_L^2}}.
\end{align}

\noindent  This state shows features in common with more conventional adiabatic passage evolution.  As $\Omega_R(0) \gg \Omega_L(0)$, we retrieve the starting configuration, i.e. $\ket{D_0} \rightarrow \ket{1\bar{1}\bar{1}\bar{1}\bar{1}}$.  In the adiabatic limit, the state of $M$ is always $\bar{1}$, and the final state has excitation only in the leaf nodes.   The final state is, however, more complicated in the two-excitation subspace.  We observe the formation of a superposition of a  W-like state: $\ket{\bar{1}\bar{1}1\bar{1}} + \ket{\bar{1}\bar{1}\bar{1}1}$, and the state $\ket{\bar{1}\bar{1}00}$, with most of the population in $\ket{\bar{1}\bar{1}00}$.  We find that $\ket{D_0}$ has less entanglement than a W state (as measured by entanglement of formation), and less entanglement than the state formed from the one-excitation state. 



The three and four leaf node configurations with two-excitations are more complicated, and we do not show the null states, as these are complicated and do not provide additional insight into the evolution.  The final state configuration for the three leaf node configuration gives rise to a superposition of two W-like states:

\begin{align}
\ket{D_0(t_{\max})} = \frac{2}{\sqrt{15}}\left(\ket{\bar{1}\bar{1}\bar{1}00} + \ket{\bar{1}\bar{1}0\bar{1}0} + \ket{\bar{1}\bar{1}00\bar{1}}\right) + \frac{1}{\sqrt{15}}\left(\ket{\bar{1}\bar{1}1\bar{1}\bar{1}} + \ket{\bar{1}\bar{1}\bar{1}1\bar{1}} + \ket{\bar{1}\bar{1}\bar{1}\bar{1}1}\right).
\end{align}
\noindent  This entangled state includes all of the possible ways that the two excitations can be shared between the leaf nodes, but the system is more strongly weighted towards configurations where the excitations are shared most evenly.  We term this sharing the \textit{egalitarian principle}, i.e. we expect the adiabatic passage to always favour the configurations where the excitations are most equally shared between the leaf nodes. 

Similarly, the four leaf node configuration with two excitations also obeys the egalitarian principle, with a contribution from a W-like state, and states where the excitations are shared on two of the leaf nodes with an amplitude twice that of the W-like state.  

\begin{align}
\ket{D_0(t)_{\max}} = & \frac{1}{\sqrt{7}} (\ket{\bar{1}\bar{1}00\bar{1}\bar{1}} + \ket{\bar{1}\bar{1}0\bar{1}0\bar{1}} + \ket{\bar{1}\bar{1}0\bar{1}\bar{1}0} + \ket{\bar{1}\bar{1}\bar{1}00\bar{1}} + \ket{\bar{1}\bar{1}\bar{1}0\bar{1}0} + \ket{\bar{1}\bar{1}\bar{1}\bar{1}00})   \nonumber \\ 
& + \frac{1}{2\sqrt{7}} \left( \ket{\bar{1}\bar{1}1\bar{1}\bar{1}\bar{1}} + \ket{\bar{1}\bar{1}\bar{1}1\bar{1}\bar{1}} + \ket{\bar{1}\bar{1}\bar{1}\bar{1}1\bar{1}} + \ket{\bar{1}\bar{1}\bar{1}\bar{1}\bar{1}1} \right).
\end{align}
\noindent
As before, the system is weighted towards configurations where the excitations are shared as evenly as possible, and this results in a state with less entanglement than configurations with fewer leaf nodes.

\section{Spin-three-half networks}
\begin{figure}[tb]
\centering
    \includegraphics[width=\linewidth]{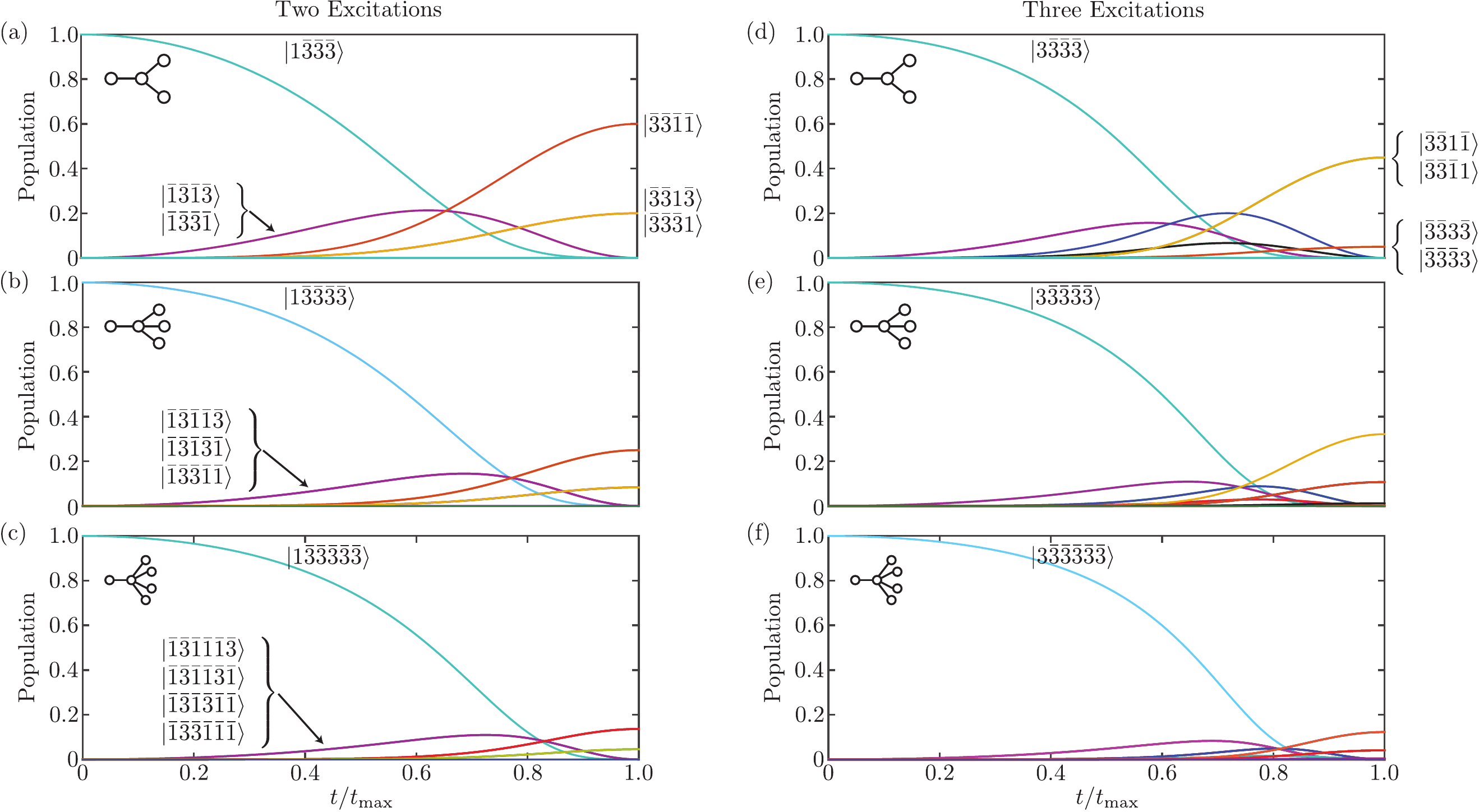}
    \caption{Evolution of systems of spin-three-half particles with two and three excitations.  The two excitation results for (a) two leaf nodes, (b) three leaf nodes and (c) four leaf nodes, are qualitatively similar to the analogous spin-one systems, although quantitively different due to the different couplings.  Three excitations for (d) two leaf nodes, (e) three leaf nodes, and (f) four leaf nodes, exhibit excitation sharing following the egalitarian principle as before, albeit with increased complexity due to the large number of states involved in the transport.  The states typically not shown here for reasons of space, but all of the final states can be found in Table~\ref{table:nonlin}.}
    \label{fig spin three halves}
\end{figure}

We now consider spin-three-half networks.  We again define the eigenstates with respect to the $J_Z$ projection, labelling the states $\ket{3}, \ket{1}, \ket{\bar{1}}, \ket{\bar 3}$ corresponding to the eigenvalues $+3\hbar/2$,  $+\hbar/2$,  $-\hbar/2$, $-3\hbar/2$ respectively. 
In the spin-three-half system, transport can be achieved by using one, two or three excitations. When limiting the system to one excitation, the transport achieved in all configurations is the equivalent of the spin-half case and the entanglement of formation was found to be equal in both cases. When limiting the system to two excitations, Figs.~\ref{fig spin three halves} (a), (b) and (c), the transport produces final states that are equivalent to those obtained in the spin-one case.  However the final states occur with different populations here than they do in the spin-one system. This is due to the additional $\sqrt{3}$ factor from the $J^+$ and $J^-$ spin operators, which also modifies the degree of entanglement produced.  Such results are expected to quantitatively affect analogous transport through higher spin networks (i.e. networks with spin $>3/2$) without modifying the qualitative form of the adiabatic passage or the final entangled states. 

With three excitations (initial state $\ket{3\bar{3}\bar{3}\cdots\bar{3}}$) the evolution is more complicated still, as shown in Figs.~\ref{fig spin three halves} (d), (e) and (f).  However we still find that the egalitarian principle holds, with all possible final states represented in the final superposition, but the states with the most equal sharing of the excitation favoured.  This sharing of excitations reduces the overall entanglement in systems with increasing number of leaf nodes.

\section{Discussion and conclusions}

Dark state adiabatic passage can be used to generate entanglement in small networks, and we have shown the properties of the evolution and entanglement in networks of spin-half, spin-one and spin-three-half systems.  Our results for the final state and entanglement of formation for these networks are summarised in the Table~\ref{table:nonlin}.  Networks of spin-half systems show the creation of W-like states across the leaf nodes, and can be understood analogously to particle splitting in spatial adiabatic passage networks \cite{HGH2011,RV2012}.  However the higher spin systems exhibit what we term an egalitarian principle, where the excitations are distributed across all possible leaf nodes, but with the states with the most equal distribution of the excitations more strongly represented in the final entangled state.  This overrepresentation of states with equal sharing reduces the entanglement of formation observed in the high spin systems, with the consequence that systems with fewer excitations and fewer leaf nodes exhibit greater entanglement.  This result is likely to be significant in the development of quantum protocols for high-spin, qudit systems.


\begin{table}[tb]
\caption{Summary of final states following multiple recipient dark state adiabatic passage, and entanglement of formation.}
\centering
\begin{tabular}{ccc l l l}
\hline\hline
Figure&Spin&Nodes&$\Psi_0$&$\Psi_f$&Entanglement \T\B \\ [0.5ex] 
\hline
2a&1/2&2&$\ket{1\bar{1}\bar{1}\bar{1}}$&$\frac{1}{\sqrt{2}}\left(\ket{\bar{1}\bar{1}\bar{1}1} + \ket{\bar{1}\bar{1}1\bar{1}}\right)$&1\T\B\\
\hline
2b&1/2&3&$\ket{1\bar{1}\bar{1}\bar{1}\bar{1}}$&$\frac{1}{\sqrt{3}}\left(\ket{\bar{1}\bar{1}\bar{1}\bar{1}1} + \ket{\bar{1}\bar{1}\bar{1}1\bar{1}} + \ket{\bar{1}\bar{1}1\bar{1}\bar{1}}\right)$&0.9183\T\B\\
\hline
2c&1/2&4&$\ket{1\bar{1}\bar{1}\bar{1}\bar{1}\bar{1}}$&$\frac{1}{2}\left(\ket{\bar{1}\bar{1}1\bar{1}\bar{1}\bar{1}} + \ket{\bar{1}\bar{1}\bar{1}1\bar{1}\bar{1}} + \ket{\bar{1}\bar{1}\bar{1}\bar{1}1\bar{1}} + \ket{\bar{1}\bar{1}\bar{1}\bar{1}\bar{1}1}\right)$&0.8113\T\B\\
\hline
3a&1&2&$\ket{0\bar{1}\bar{1}\bar{1}}$&$\frac{1}{\sqrt{2}}\left(\ket{\bar{1}\bar{1}0\bar{1}} + \ket{\bar{1}\bar{1}\bar{1}0}\right)$&1\T\\
\hline
3b&1&3&$\ket{0\bar{1}\bar{1}\bar{1}\bar{1}}$&$\frac{1}{\sqrt{3}}\left(\ket{\bar{1}\bar{1}0\bar{1}\bar{1}} + \ket{\bar{1}\bar{1}\bar{1}0\bar{1}} + \ket{\bar{1}\bar{1}\bar{1}\bar{1}0}\right)$&0.9183\T\B\\
\hline
3c&1&4&$\ket{0\bar{1}\bar{1}\bar{1}\bar{1}\bar{1}}$&$\frac{1}{\sqrt{4}}\left(\ket{\bar{1}\bar{1}0\bar{1}\bar{1}\bar{1}} + \ket{\bar{1}\bar{1}\bar{1}0\bar{1}\bar{1}} + \ket{\bar{1}\bar{1}\bar{1}\bar{1}0\bar{1}} + \ket{\bar{1}\bar{1}\bar{1}\bar{1}\bar{1}0}\right)$&0.8113\T\B\\
\hline
3d&1&2&$\ket{1\bar{1}\bar{1}\bar{1}}$&$\sqrt{\frac{2}{3}}\left(\ket{\bar{1}\bar{1}00}\right) + \sqrt{\frac{1}{6}}\left(\ket{\bar{1}\bar{1}\bar{1}1} + \ket{\bar{1}\bar{1}1\bar{1}}\right)$&0.8072\Tt\B\\
\hline
3e&1&3&$\ket{1\bar{1}\bar{1}\bar{1}\bar{1}}$&\pbox{20cm}{$\sqrt{\frac{4}{15}}\left(\ket{\bar{1}\bar{1}\bar{1}00} + \ket{\bar{1}\bar{1}0\bar{1}0} + \ket{\bar{1}\bar{1}00\bar{1}}\right)$ \\ $+ \sqrt{\frac{1}{15}}\left(\ket{\bar{1}\bar{1}1\bar{1}\bar{1}} + \ket{\bar{1}\bar{1}\bar{1}1\bar{1}} + \ket{\bar{1}\bar{1}\bar{1}\bar{1}1}\right)$}&0.7264\Ttt\B\\
\hline
3f&1&4&$\ket{1\bar{1}\bar{1}\bar{1}\bar{1}\bar{1}}$&\pbox{20cm}{ $\frac{1}{\sqrt{7}} (\ket{\bar{1}\bar{1}00\bar{1}\bar{1}} + \ket{\bar{1}\bar{1}0\bar{1}0\bar{1}} + \ket{\bar{1}\bar{1}0\bar{1}\bar{1}0}$ \\ $+ \ket{\bar{1}\bar{1}\bar{1}00\bar{1}} + \ket{\bar{1}\bar{1}\bar{1}0\bar{1}0} + \ket{\bar{1}\bar{1}\bar{1}\bar{1}00})$ \\$+ \frac{1}{\sqrt{28}} \left( \ket{\bar{1}\bar{1}1\bar{1}\bar{1}\bar{1}} + \ket{\bar{1}\bar{1}\bar{1}1\bar{1}\bar{1}} + \ket{\bar{1}\bar{1}\bar{1}\bar{1}1\bar{1}} + \ket{\bar{1}\bar{1}\bar{1}\bar{1}\bar{1}1} \right)$ }& 0.5152\Ttt\\
\hline
-&3/2&2&$\ket{\bar{1}\bar{3}\bar{3}\bar{3}}$&$\frac{1}{\sqrt{2}} \left(\ket{\bar{3}\bar{3}\bar{1}\bar{3}} + \ket{\bar{3}\bar{3}\bar{3}\bar{1}}\right)$&1\T\B\\
\hline
-&3/2&3&$\ket{\bar{1}\bar{3}\bar{3}\bar{3}\bar{3}}$&$\frac{1}{\sqrt{3}}\left(\ket{\bar{3}\bar{3}\bar{1}\bar{3}\bar{3}} + \ket{\bar{3}\bar{3}\bar{3}\bar{1}\bar{3}} + \ket{\bar{3}\bar{3}\bar{3}\bar{3}\bar{1}}\right)$&0.9183\T\B\\
\hline
-&3/2&4&$\ket{\bar{1}\bar{3}\bar{3}\bar{3}\bar{3}\bar{3}}$&$\frac{1}{2}\left(\ket{\bar{3}\bar{3}\bar{1}\bar{3}\bar{3}\bar{3}} + \ket{\bar{3}\bar{3}\bar{3}\bar{1}\bar{3}\bar{3}} + \ket{\bar{3}\bar{3}\bar{3}\bar{3}\bar{1}\bar{3}} + \ket{\bar{3}\bar{3}\bar{3}\bar{3}\bar{3}\bar{1}}\right)$&0.8113\T\B\\
\hline
4a&3/2&2&$\ket{1\bar{3}\bar{3}\bar{3}}$&$\sqrt{\frac{3}{5}}\left(\ket{\bar{3}\bar{3}\bar{1}\bar{1}}\right) +\sqrt{ \frac{1}{5}}\left(\ket{\bar{3}\bar{3}1\bar{3}} + \ket{\bar{3}\bar{3}\bar{3}1}\right)$&0.9021\T\B\\
\hline
4b&3/2&3&$\ket{1\bar{3}\bar{3}\bar{3}\bar{3}}$&\pbox{20cm}{$\frac{1}{2}\left(\ket{\bar{3}\bar{3}\bar{1}\bar{1}\bar{3}} + \ket{\bar{3}\bar{3}\bar{1}\bar{3}\bar{1}} + \ket{\bar{3}\bar{3}\bar{3}\bar{1}\bar{1}}\right)$\\$+ \frac{1}{\sqrt{12}}\left(\ket{\bar{3}\bar{3}1\bar{3}\bar{3}} + \ket{\bar{3}\bar{3}\bar{3}1\bar{3}} + \ket{\bar{3}\bar{3}\bar{3}\bar{3}1}\right)$}&0.7080\Ttt\B\\
\hline
4c&3/2&4&$\ket{1\bar{3}\bar{3}\bar{3}\bar{3}\bar{3}}$&\pbox{20cm}{$\sqrt{\frac{3}{22}} (\ket{\bar{3}\bar{3}\bar{1}\bar{1}\bar{3}\bar{3}} + \ket{\bar{3}\bar{3}\bar{1}\bar{3}\bar{1}\bar{3}} + \ket{\bar{3}\bar{3}\bar{3}\bar{3}\bar{1}\bar{1}} $\\$+ \ket{\bar{3}\bar{3}\bar{1}\bar{3}\bar{3}\bar{1}} + \ket{\bar{3}\bar{3}\bar{3}\bar{1}\bar{1}\bar{3}} 
+ \ket{\bar{3}\bar{3}\bar{3}\bar{1}\bar{3}\bar{1}})$\\$ + \frac{1}{\sqrt{22}} (\ket{\bar{3}\bar{3}1\bar{3}\bar{3}\bar{3}} + \ket{\bar{3}\bar{3}\bar{3}1\bar{3}\bar{3}} + \ket{\bar{3}\bar{3}\bar{3}\bar{3}1\bar{3}} + \ket{\bar{3}\bar{3}\bar{3}\bar{3}\bar{3}1})$}&0.4997\Ttt\B\\
\hline
4d&3/2&2&$\ket{3\bar{3}\bar{3}\bar{3}}$&$\frac{3}{2\sqrt{5}}\left(\ket{\bar{3}\bar{3}1\bar{1}} + \ket{\bar{3}\bar{3}\bar{1}1}\right) + \frac{1}{2\sqrt{5}}\left(\ket{\bar{3}\bar{3}3\bar{3}} + \ket{\bar{3}\bar{3}\bar{3}3}\right)$&0.9763\Tt\B\\
\hline
4e&3/2&3&$\ket{3\bar{3}\bar{3}\bar{3}\bar{3}}$&\pbox{20cm}{$\frac{3\sqrt{3}}{\sqrt{85}} \ket{\bar{3}\bar{3}\bar{1}\bar{1}\bar{1}} + \frac{3}{\sqrt{85}}(\ket{\bar{3}\bar{3}\bar{3}\bar{1}1} + \ket{\bar{3}\bar{3}1\bar{1}\bar{3}}$\\$ + \ket{\bar{3}\bar{3}1\bar{3}\bar{1}} + \ket{\bar{3}\bar{3}\bar{1}1\bar{3}} + \ket{\bar{3}\bar{3}\bar{1}\bar{3}1} + \ket{\bar{3}\bar{3}\bar{3}1\bar{1}})$\\ $ + \frac{1}{\sqrt{85}}(\ket{\bar{3}\bar{3}3\bar{3}\bar{3}} + \ket{\bar{3}\bar{3}\bar{3}3\bar{3}} + \ket{\bar{3}\bar{3}\bar{3}\bar{3}3})$}&0.6297\Ttt\\
\hline
4f&3/2&4&$\ket{3\bar{3}\bar{3}\bar{3}\bar{3}\bar{3}}$&\pbox{20cm}{$ \frac{3\sqrt{3}}{\sqrt{220}} (\ket{\bar{3}\bar{3}\bar{1}\bar{1}\bar{1}\bar{3}} + \ket{\bar{3}\bar{3}\bar{1}\bar{1}\bar{3}\bar{1}} + $\\$\ket{\bar{3}\bar{3}\bar{1}\bar{3}\bar{1}\bar{1}} + \ket{\bar{3}\bar{3}\bar{3}\bar{1}\bar{1}\bar{1}}) + \frac{3}{\sqrt{220}} (\ket{\bar{3}\bar{3}1\bar{1}\bar{3}\bar{3}} $\\$+ \ket{\bar{3}\bar{3}1\bar{3}\bar{1}\bar{3}} + \ket{\bar{3}\bar{3}1\bar{3}\bar{3}\bar{1}} + \ket{\bar{3}\bar{3}\bar{1}1\bar{3}\bar{3}} $\\$+ \ket{\bar{3}\bar{3}\bar{1}\bar{3}1\bar{3}} + \ket{\bar{3}\bar{3}\bar{1}\bar{3}\bar{3}1} + \ket{\bar{3}\bar{3}\bar{3}1\bar{1}\bar{3}} $\\$+ \ket{\bar{3}\bar{3}\bar{3}1\bar{3}\bar{1}} + \ket{\bar{3}\bar{3}\bar{3}\bar{1}1\bar{3}} + \ket{\bar{3}\bar{3}\bar{3}\bar{1}\bar{3}1} $\\$+ \ket{\bar{3}\bar{3}\bar{3}\bar{3}1\bar{1}} + \ket{\bar{3}\bar{3}\bar{3}\bar{3}\bar{1}1}) + \frac{1}{\sqrt{220}} (\ket{\bar{3}\bar{3}3\bar{3}\bar{3}\bar{3}} $\\$+ \ket{\bar{3}\bar{3}\bar{3}3\bar{3}\bar{3}} + \ket{\bar{3}\bar{3}\bar{3}\bar{3}3\bar{3}} + \ket{\bar{3}\bar{3}\bar{3}\bar{3}\bar{3}3})$}&0.3889\Tttt\B\\
\hline

\end{tabular}
\label{table:nonlin}
\end{table}

\section*{Acknowledgements}
ADG acknowledges the ARC for financial support (DP130104381).  



\begin{thebibliography}{99}
\bibitem{KTS2007} P. Kr\'{a}l, I. Thanopulos, M. Shapiro, Rev. Mod. Phys. \textbf{79}, 53 (2007).

\bibitem{GRB+1988} U. Gaubatz, P. Rudecki, M. Becker, S. Schiemann, M. K\"{u}lz and K. Bergmann, Chem. Phys. Lett. \textbf{149}, 463 (1988)

\bibitem{PYH2005} T. Peters, L. P. Yatsenko, and T. Halfmann, ÒExperimental demonstration of selective coherent population transfer via a continuum,Ó Phys. Rev. Lett. {\bf 95}, 103601 (2005).

\bibitem{DSH+2009} F. Dreisow, A. Szameit, M. Heinrich, R. Keil, S. Nolte, A. T\"{u}nnermann, and S. Longhi, ÒAdiabatic transfer of light via a continuum in optical waveguides,Ó Opt. Lett. {\bf 34}, 2405Ð2407 (2009). 

\bibitem{BV2002} T. Brandes and T. Vorrath, Phys. Rev. B \textbf{66}, 075341 (2002).

\bibitem{ECB+2004} K. Eckert, M. Lewenstein, R. Corbal\'{a}n, G. Birkl, W. Ertmer, and J. Mompart, Phys. Rev. A {\bf 70}, 023606 (2004).

\bibitem{SB2004} J. Siewert and T. Brandes, Adv. Solid State Phys. {\bf 44}, 181 (2004).

\bibitem{GCH+2004} A. D. Greentree, J. H. Cole, A. R. Hamilton, and L. C. L. Hollenberg, Phys. Rev. B, {\bf 70}, 235317 (2004).

\bibitem{GKW2006} E. M. Graefe, H. J. Korsch, and D. Witthaut, Phys. Rev. A {\bf 73}, 013617 (2006).

\bibitem{RCP+2008} M. Rab, J. H. Cole, N. G. Parker, A. D. Greentree, L. C. L. Hollenberg, and A. M. Martin, Phys. Rev. A {\bf 77}, 061602(R) (2008).













\bibitem{P2006} E. Paspalakis, Opt. Commun. {\bf 258}, 30 (2006).

\bibitem{LDO+2007} S. Longhi, G. Della Valle, M. Ornigotti, and P. Laporta, Phys. Rev. B {\bf 76}, 201101(R) (2007).

\bibitem{BRG+2012} C. J. Bradly, M. Rab, A. D. Greentree, and A. M. Martin, Phys. Rev. A {\bf 85}, 053609 (2012).

\bibitem{MMA2014} R. Menchon-Enrich, J. Mompart, and V. Ahufinger, Phys. Rev. B {\bf 89}, 094304 (2014).

\bibitem{DRK2015} E. Dimova, A. Rangelov, and E. Kyoseva, arXiv:1502.04857 (2015).

\bibitem{OEO+2007} T. Ohshima, A. Ekert, D. K. L. Oi, D. Kaslizowski, and L. C. Kwek, e-print arXiv:quant-ph/0702019.

\bibitem{OSF+2013} S. Oh, Y.-P. Shim, J. Fei, M. Friesen, and X. Hu, Phys. Rev. A \textbf{87}, 022332 (2013).

\bibitem{GK2014} A. D. Greentree and B. Koiller, Phys. Rev. A \textbf{90}, 012319 (2014)

\bibitem{USB1999} R. G. Unanyan, B. W. Shore, and K. Bergmann, Phys. Rev. A {\bf 59} 2910 (1999).

\bibitem{KR2002} Z. Kis and F. Renzoni, Phys. Rev. A {\bf 65}, 032318 (2002).

\bibitem{DGH2007} S. J. Devitt, A. D. Greentree, and L. C. L. Hollenberg, Quantum Information Processing {\bf 6}, 229 (2007).

\bibitem{HNM+2015} A. P. Hope, T. G. Nguyen, A. Mitchell and A. D. Greentree J. Phys. B {\bf 48}, 055503 (2015),

\bibitem{JG2010} L. M. Jong and A. D. Greentree, Phys. Rev. B {\bf 81}, 035311 (2010).

\bibitem{HGH2011} C. D. Hill, A. D. Greentree, and L. C. L. Hollenberg, New J. Phys. {\bf 13}, 125002 (2011).

\bibitem{DOS+2009} F. Dreisow, M. Ornigotti, A. Szameit, M. Heinrich, R. Keil, S. Nolte, A. Tunnermann, and S. Longhi, Appl. Phys. Lett. {\bf 95}, 261102 (2009).

\bibitem{CCR+2012} C. Ciret, V. Coda, A. A Rangelov, D. N Neshev, and G. Montemezzani, Opt. Lett. {\bf 37}, 3789 (2012).

\bibitem{CKR+2012} K. Chung, T. J. Karle, M. Rab, A. D. Greentree, and S. Tomljenovic-Hanic, Opt. Exp. {\bf 20}, 23108 (2012).

\bibitem{L2014} S. Longhi Coherent transfer by adiabatic passage in two-dimensional lattices Ann. Phys., NY {\bf 348}, 161 (2014).

\bibitem{GDH2006} A. D. Greentree, S. J. Devitt, and L. C. L. Hollenberg, Phys. Rev. A {\bf 73}, 032319 (2006).

\bibitem{RV2012} A. A. Rangelov and N. V. Vitanov, Phys. Rev. A {\bf 85}, 055803 (2012).

\bibitem{W2001} W. K. Wooters, Quantum Inf. and Computation {\bf 1}, 27 (2001).

\bibitem{WC2001} A. Wong and N. Christensen, Phys. Rev. A {\bf 63}, 044301 (2001). 



\bibitem{H83} F. D. M. Haldane, Phys. Rev. Lett. 50, 1153 (1983).

\bibitem{AKL+87} I. Affleck, T. Kennedy, E. H. Lieb, and H. Tasaki, Phys. Rev. Lett. 59, 799 (1987).

\end{thebibliography}
\end{document}